\documentclass[preprint,12pt, a4paper]{elsarticle}

\usepackage{amssymb}
\usepackage{hyperref}
\usepackage{listings}
\usepackage{xcolor}
\usepackage[version=4]{mhchem}
\usepackage{todonotes}
\journal{SoftwareX}

\begin{document}
\renewcommand{\labelenumii}{\arabic{enumi}.\arabic{enumii}}

\begin{frontmatter}



\title{\textsc{MatNexus}: A Comprehensive Text Mining and Analysis Suite for Materials Discovery}


\author[label1]{L. Zhang\corref{cor1}}
\author[label1]{M. Stricker}
\address[label1]{Interdisciplinary Centre for Advanced Materials Simulation, Ruhr-Universität Bochum, 44801 Bochum, \{Lei.Zhang-w2i, markus.stricker\}@rub.de}
\cortext[cor1]{Corresponding author.}

\begin{abstract}

\textsc{MatNexus} is a specialized software for the automated collection, processing, and analysis of the text from scientific articles. Through an integrated suite of modules, the \textsc{MatNexus} facilitates the retrieval of scientific articles, processes textual data for insights, generates vector representations suitable for machine learning, and offers visualization capabilities for word embeddings. With the vast volume of scientific publications, \textsc{MatNexus} stands out as an end-to-end tool for researchers aiming to gain insights from scientific literature in material science, making the exploration of materials, such as the electrocatalyst examples we show here, efficient and insightful.
\end{abstract}

\begin{graphicalabstract}
\includegraphics[width=\textwidth]{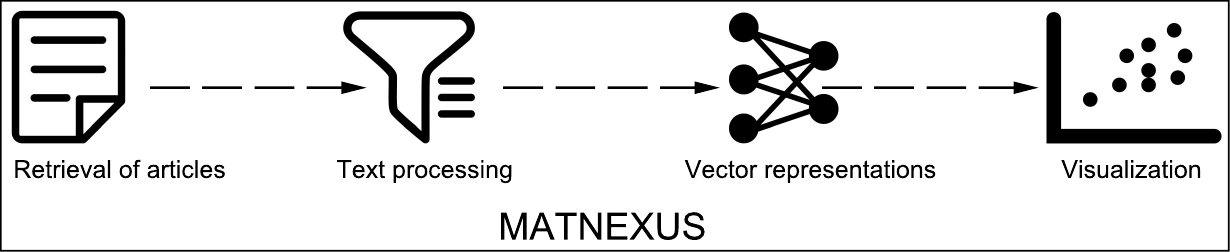}
\end{graphicalabstract}

\begin{highlights}
\item \textsc{MatNexus} offers an integrated suite for automated collection, processing, and analysis of scientific articles in material science.
\item Features include article retrieval, insights extraction from textual data, and generating machine-learning ready vector representations.
\item Facilitates efficient exploration and visualization of materials, exemplified with electrocatalyst insights.
\end{highlights}


\end{frontmatter}
\clearpage

\section*{Metadata}
\label{}

\begin{table}[!h]
\caption{Code metadata.}
\begin{tabular}{|l|p{6.5cm}|p{6.5cm}|}
\hline
\textbf{Nr.} & \textbf{Code metadata description} \\
\hline
C1 & Current code version & v1.0 \\
\hline
C2 & Permanent link to code/repository used for this code version & \url{https://github.com/lab-mids/matnexus} \\
\hline
C3  & Permanent link to Reproducible Capsule & 'none'\\
\hline
C4 & Legal Code License   & GNU LGPLv3.0\\
\hline
C5 & Code versioning system used & git \\
\hline
C6 & Software code languages, tools, and services used & Python \\
\hline
C7 & Compilation requirements, operating environments \& dependencies & MS windows, Linux \\
\hline
C8 & If available Link to developer documentation/manual &  \url{https://github.com/lab-mids/matnexus} \\
\hline
C9 & Support email for questions &  Lei.Zhang-w2i@rub.de\\
\hline
\end{tabular}
\label{codeMetadata} 
\end{table}

\section{Motivation and Significance}

The rapid development and dynamism of materials science is evidenced by the growing number of academic publications~\cite{Kademani20131275}.
Many discoveries and methodologies have given rise to vast amount of data in form text published in scientific articles that researchers have to grapple with~\cite{Mahdi2020119554}.
This surge in knowledge published in the form of (very) many scientific articles, however, also introduces a problem: it is becoming increasingly more difficult for individual researchers to keep track and make use of the available knowledge and identify trends in research~\cite{Kilicoglu20171400}.
Consequently, a new class of tools is required that can query, process, and visualize the knowledge existing in literature.
\textsc{\textsc{MatNexus}} is exactly solving this problem: efficient extraction and aggregation of knowledge for insightful, actionable research findings based on scientific literature and provide output suitable for machine learning applications aiding in materials discovery.

\textsc{MatNexus} not only offers a functional advantage but also represents a step forward in innovation, bridging the gap between extensive academic datasets and economically usable research outcomes.
Because individual results, i.e. individual articles, from academic research are largely selective and incomplete.
The whole body of scientific literature, however, allows the extraction of valuable knowledge that is less selective and as complete as possible provided sufficient access to literature data.
\textsc{MatNexus}' design focuses on empowering users, ensuring that researchers play an active role in the data analysis and interpretation process.
The toolset comprises a \texttt{PaperCollector} to efficiently collect abstracts from scientific literature through the Scopus API~\cite{ScopusAPI}\footnote{This is the current version but any API which provides abtracts can be used.} using pybliometrics~\cite{Rose2019}, a \texttt{TextProcessor} for filtering operations, as well as a \texttt{VecGenerator} to create embeddings~\cite{vRehruvrek2010} and \texttt{VecVisualizer} for in-depth analysis and visualization.

\textsc{MatNexus}' potential impact on the materials science community is significant. In a world of rapidly changing scientific paradigms, the need for efficient and targeted research tools is ever-growing~\cite{Tian2019}.
By providing a solution for literature processing and analysis, it aims to simplify existing research methods, enhancing both the pace and precision of discoveries~\cite{Maffettone2021}. More importantly, it is designed to highlight new opportunities for materials discovery and property prediction, and thereby enabling accelerated advancements in the field~\cite{Tshitoyan2019}.

\textsc{MatNexus} is conceptually built around a workflow.
It starts with a query, using the \texttt{PaperCollector}.
This module acts as the foundation, which provides text data as collected through the Scopus API~\cite{Rose2019}.
The resulting data set is subsequently refined by the \texttt{TextProcessor}, details of this are presented later.
Subsequently, the \texttt{VecGenerator} transforms the textual data into word embeddings~\cite{mikolov2013word2vec} and thereby providing a basis for machine learning applications.
Lastly, the \texttt{VecVisualizer} can be used as an interface for visualization\cite{plotly,Hunter:2007,harris2020array} including dimensionality reduction of the word embeddings for intuitive exploration of the data relationships\cite{scikit-learn}, patterns, and anomalies.

Comparative analysis reveals that while there are a host of scientific literature analysis tools~\cite{Beasley2018}, like PubMed catering to niches such as biomedical research~\cite{PubMed,Williamson201916}, the landscape is devoid of specialized tools for materials science which provides a full stack from search to visualization capabilities.
Established methods, such as Term Frequency-Inverse Document Frequency (TF-IDF)~\cite{jones1972statistical} and Word2Vec~\cite{mikolov2013word2vec}, have been instrumental in general computer-aided text processing~\cite{Li2018450,Arora2021199,Zhang2020199}. However, \textsc{MatNexus} distinguishes itself with its dedicated focus on materials science, particularly for materials discovery.
The complete workflow is presented here with materials discovery in mind: The concept of a material as well as material properties are an integral part of the design.
Further, \textsc{MatNexus} is designed as a versatile tool and can, in principle, be appropriated for other research domains. 
We present its utility in electrocatalysis due to the intricacies involved in discovering new High-Entropy Alloys (HEA) electrocatalysts.
With this we demonstrate our tool's capability in handling complex research scenarios.

\section{Software description}
\subsection{Software architecture}

\begin{figure}[!hbt]
    \centering
    \includegraphics[width=0.75\linewidth]{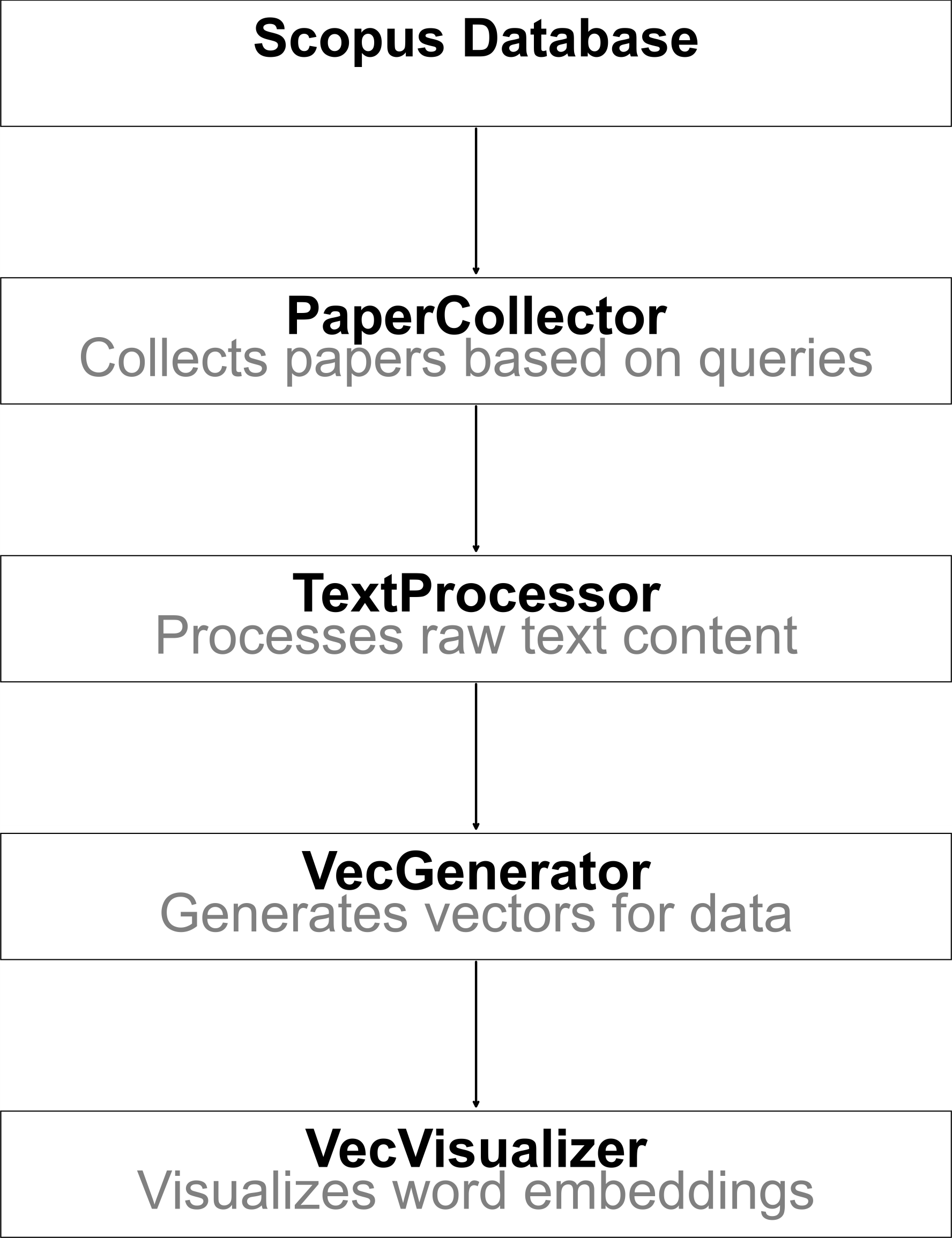}
    \caption{Data flow chart of \textsc{MatNexus} modules.}
    \label{fig:flow chart}
\end{figure}

\textsc{MatNexus} has a modular architecture, ensuring ease of use, scalability, maintainability, and extendability. The primary modules are \texttt{PaperCollector}, \texttt{TextProcessor}, \texttt{VecGenerator}, and \texttt{VecVisualizer}, as illustrated in Figure~\ref{fig:flow chart}.

\subsection{Software Functionalities}

\textsc{MatNexus} provides functionality for a series of tasks that streamline a text data-based research process in materials science:

\begin{itemize}
    \item \textbf{Automated Literature Collection}: Utilizing the \texttt{PaperCollector} module, \textsc{MatNexus} enables researchers to tailor search queries based on specific parameters like keywords, publication dates, and open access status. This module currently interfaces with the Scopus database using the pybliometrics module~\cite{Rose2019}. While it primarily fetches academic articles from Scopus, focusing on their abstracts, the underlying architecture of \textsc{MatNexus} allows addition of other databases or APIs. Collected abstracts are saved for further processing and analysis\cite{reback2020pandas}.
    \item \textbf{Structured Textual Processing}: Following the collection of data, the \texttt{TextProcessor} module is employed to curate the raw textual data into a structured and coherent format\cite{harris2020array,bird2009natural}. This module provides several functions to refine the text data for the following word embeddings and analysis:
    \begin{itemize}
        \item \textit{Sentence Filtering}: Certain sentences are identified and filtered out to get rid of text data unrelated to materials. Specifically, sentences ending with symbols like ©, or phrases like "\& Co." are modified to ensure proper sentence boundaries. Further, sentences containing phrases such as "©" or "rights reserved" are completely removed as they don't contribute to the material science value of the data.
        \item \textit{Word Filtering}: The text is tokenized into individual words. After that, various filters come into play. These filters are designed to exclude elements like punctuation, stop words (commonly used words that carry little meaning), and pure numeric values. Additionally, a special emphasis is placed on retaining chemical formulas while filtering out other unrelated numerical values or words.
        \item \textit{Lemmatization}: This step involves reducing words to their base or root form (e.g. only singulars instead of plurals and singulars), a process known as lemmatization. It is crucial for consolidating different forms or variations of a word into a standard form, which significantly aids in achieving a more accurate analysis because it improves entity recoginition.
       \end{itemize}
    With the completion of these processing steps, the refined data is then converted into a structured CSV (comma separated values) format. This transformation is fundamental as it ensures that the subsequent analyses are both robust and comprehensive. The structured format helps in minimizing potential inconsistencies in data interpretation by providing a clean, standardized dataset that is ready for further analysis.
    
    \item \textbf{Advanced Textual Analysis via Word2Vec Embeddings}: 
    
    The \texttt{VecGenerator} module encapsulates the analytical capabilities of \textsc{MatNexus}. Through the generation of a Word2Vec model, this module facilitates nuanced analyses, allowing researchers to discern patterns, correlations, and overarching trends within the literature~\cite{rehurek_lrec}.

    \item \textbf{Multifaceted Data Visualization}: The \texttt{VecVisualizer} module provides a suite of visualization tools, each tailored to present data in an intuitive manner. From dimensionality reduction techniques like t-SNE and isomap to material vector visualizations, this module provides graphical representations that support understanding and interpretation of complex datasets and relationships extracted from text.
    In our implementation, cosine similarity is used to measure the similarity between (word embedding) vectors. Cosine similarity quantifies the cosine of the angle between two non-zero vectors, providing a measure of similarity between them based on their orientation, irrespective of their magnitude. This measure proves to be effective in understanding the closeness, proximity or similarity between data points in a multi-dimensional space~\cite{XIA201539}. However, other similarity measurement techniques could also be employed depending on the specific requirements and nature of the data.

    \item \textbf{Material Similarity Computation and Analysis}: Beyond mere textual analysis, \textsc{MatNexus} provides the capability to compute material similarities. By comparing various materials with a reference point. This functionality offers insights into the interrelationships between different materials, potentially highlighting synergies and areas of interest for further exploration.
    
    \item \textbf{Interactive Graphical Representations}: To improve user engagement and data accessibility, \textsc{MatNexus} incorporates interactive visualizations either through similarity scatter plots or material vector representations, these interactive graphical interfaces serve as a channel for users to assess the nuances of datasets and thereby facilitating a comprehensive understanding.
\end{itemize}

\section{Illustrative Examples}

To elucidate the capabilities of \textsc{MatNexus}, we present an illustrative example that encompasses the software's major functions. In this section, our example provides an overview how \textsc{MatNexus} can be used in a Materials Discovery campaign for novel electrocatalytic materials from the class of high entropy alloys for the hydrogen evolution reaction.

For the convenience of readers and to facilitate hands-on exploration, we have provided a lightweight version of this example in the \texttt{Example} directory on our \href{https://github.com/lab-mids/matnexus/tree/main/Example}{GitHub repository}. While this GitHub example is designed for immediate experimentation and understanding, please note that it is a simplified version and might not capture the full complexity of the tool discussed in this paper.

\subsection{Literature Collection for Electrocatalyst and High Entropy Alloy}

\begin{figure}[htp!]
    \centering
    \includegraphics[width=\textwidth]{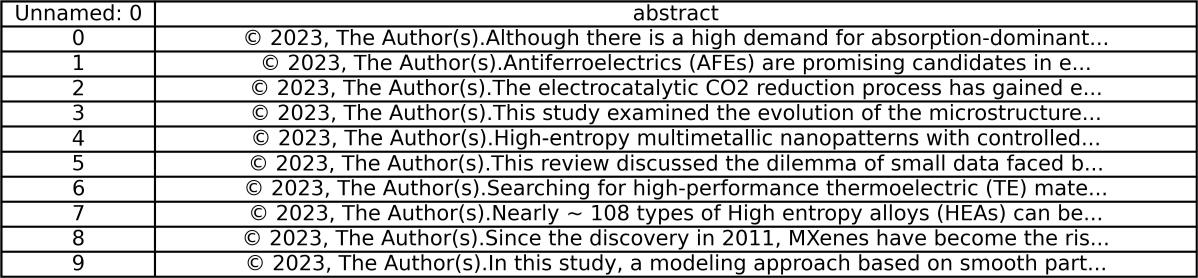}
    \caption{Screenshot of collected abstracts.}
    \label{fig:collector}
\end{figure}

Imagine a scenario where a researcher aims to gather literature on "Electrocatalyst" and "High Entropy Alloys" published until 2022. 
Utilizing the \texttt{PaperCollector} module, the researcher can specify their search parameters, and \textsc{MatNexus} will interface with the Scopus database to retrieve relevant articles. In our example, we restrict \textsc{MatNexus} to collect only Open-Access publications, thereby avoiding any copyright issue.
The result of the query is a collection of abstract from articles that fit the search criteria as illustrated in Figure~\ref{fig:collector}.

\subsection{Processing and Structuring Data}

\begin{figure}
    \centering
    \includegraphics[width=\textwidth]{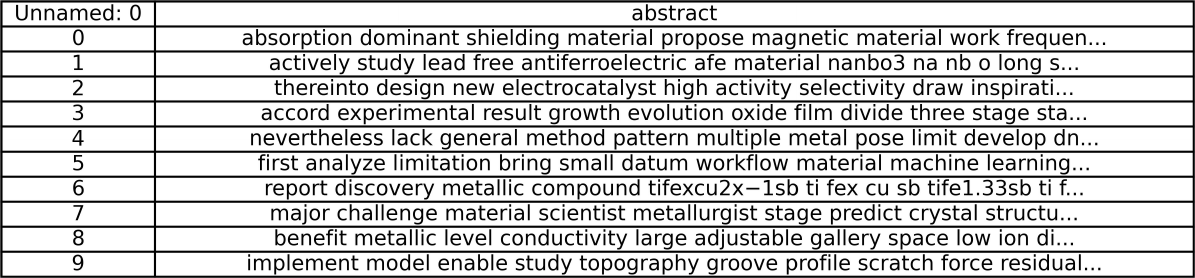}
    \caption{Screenshot of processed data}
    \label{fig:processor}
\end{figure}

After retrievel, the \texttt{TextProcessor} module is used to refine the raw abstracts. It systematically filters out irrelevant content, standardizes words through lemmatization, and retains pertinent details such as chemical formulas. The result is a set of cleaned abstracts, which are then stored in a structured CSV format (Figure~\ref{fig:processor}), ensuring a streamlined data set for subsequent analyses.

\subsection{Textual Analysis and Insight Generation}

\begin{figure}
    \centering
    \includegraphics[width=\textwidth]{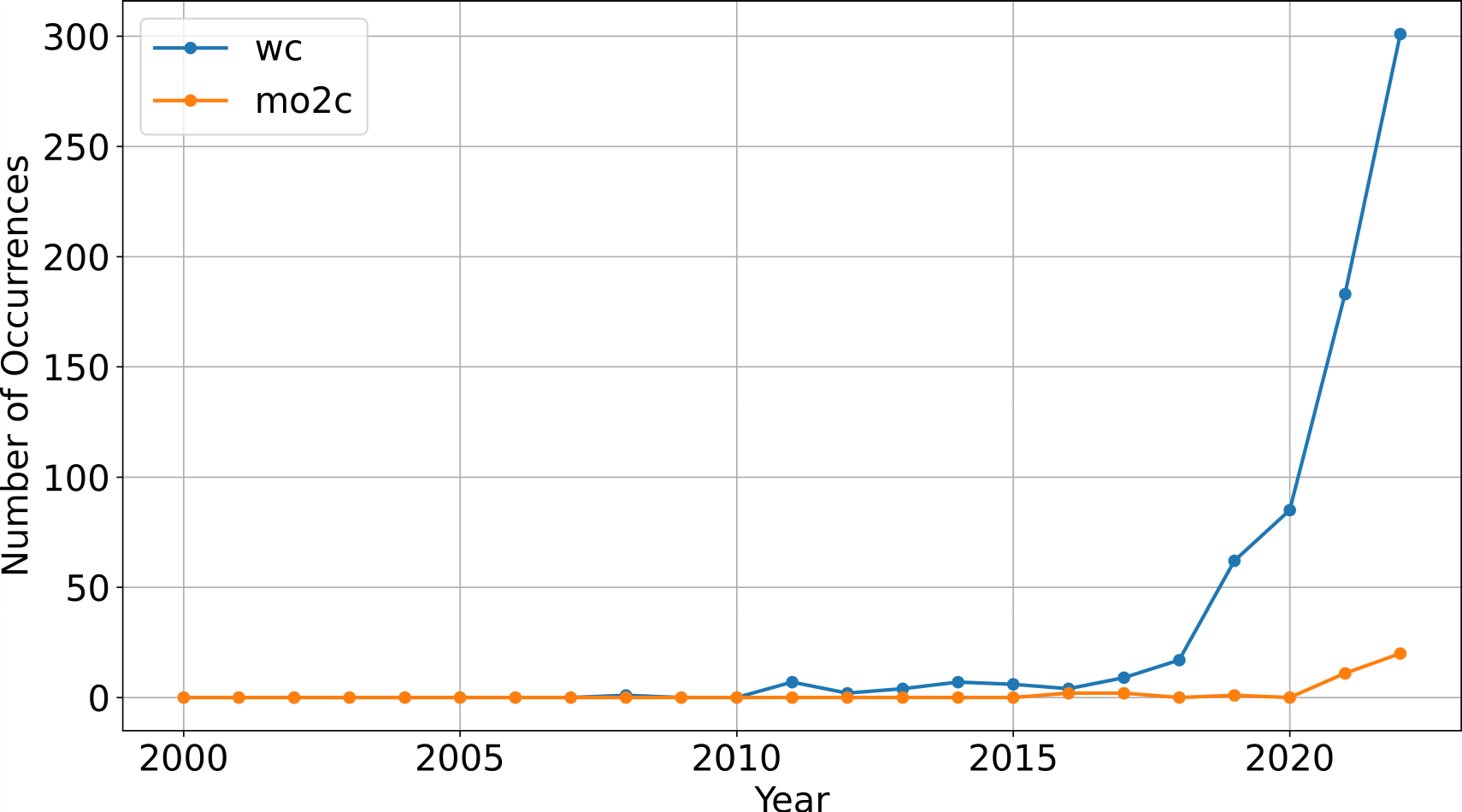}
    \caption{Term frequency trend for \texttt{'wc'} and \texttt{'mo2c'} over the years 2000 to 2022.}
    \label{fig:term_trends}
\end{figure}

\textsc{MatNexus} allows to identify frequently co-occurring terms and spot emerging areas of interest within the domain of electrocatalyts.
To illustrate this capability, consider Figure~\ref{fig:term_trends}, which shows the frequency trends for the term \texttt{'wc'} (chemical symbol for Tungsten carbide) and \texttt{'mo2c'} (Molybdenum carbide) from the years 2000 to 2022. 
From this data, it is evident that the material \texttt{'wc'} has been gradually gaining traction in the research community since 2008, with a pronounced surge from 2019 onwards. In contrast, \texttt{'mo2c'}, though less frequently mentioned than \texttt{'wc'}, has displayed a rising trend since 2016, indicating a increased interest in this material in recent years. 

The \texttt{VecGenerator} module further aids in improving understanding by extracting latent knowledge from abstracts. Using \texttt{gensim}~\cite{rehurek_lrec}, we generate a Word2Vec model from the curated data, enabling us to explore meaningful representations of terms based on their contextual similarities.

One of the primary strengths of the \texttt{VecGenerator} module is its ability to pinpoint terms that are contextually related to a given query term. For example, when the model is queried with \texttt{'pt'} (chemical symbol for Platinum), it returns: 

\begin{verbatim}
[('pd', 0.7342804074287415),
 ('rh75', 0.705005943775177),
 ('pt25', 0.6887199282646179),
 ('pt50', 0.6811999678611755),
 ('rh50', 0.6778424978256226),
 ('rh', 0.6769404411315918),
 ('ru', 0.6654356718063354),
 ('rh0.50pd0.50', 0.6543915271759033),
 ('ir48pt74ru30rh30ag74', 0.6536775231361389),
 ('ru1', 0.646355390548706)]
\end{verbatim}

Here, each tuple contains a term and its similarity score with the query term. For instance, \texttt{'pd'} (Palladium) has a high similarity score with \texttt{'pt'}, reflecting their shared applications in catalysis and their adjacent positions in the periodic table.

The terms \texttt{'rh75'}, \texttt{'pt25'}, \texttt{'pt50'}, etc., indicate specific compositions. For instance, \texttt{'rh75'} implies a composition that contains 75\% Rhodium. Such terms suggest that the associated materials frequently appear alongside or in comparison with Platinum in electrocatalysis contexts.

The presence of \texttt{'ru'} (Ruthenium) further showcases the capability to capture other metals often paired or compared with Platinum in electrocatalytic contexts. The complex term \texttt{'ir48pt74ru30rh30ag74'} suggests a multi-elemental alloy, potentially hinting at a niche or innovative area in the dataset.

This brief analysis demonstrates the depth of insights that can be gained using the \texttt{VecGenerator}, even from a simple query.
It highlights our tool's potential in guiding researchers to both well-established and emerging areas within the domain.

\subsection{Data Visualization and Interpretation}

The \texttt{VecVisualizer} module is a powerful tool for visualizing the relationships between key terms and properties within the domain of electrocatalysis. One of the most insightful visualization techniques employed is t-SNE (t-distributed Stochastic Neighbor Embedding)~\cite{LIU2021126146}. 

\begin{figure}[h]
    \centering
    \includegraphics[width=\textwidth]{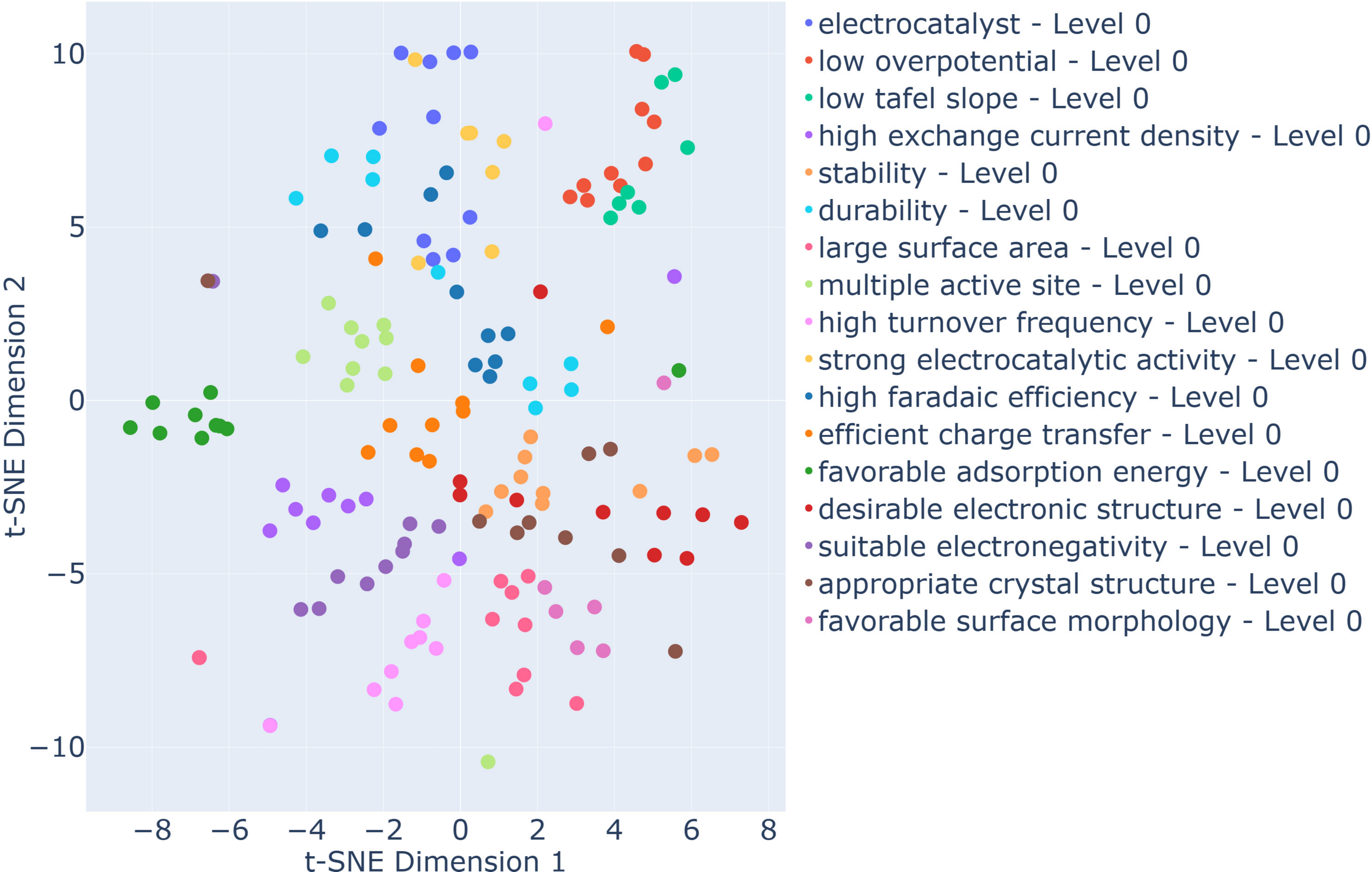}
    \caption{t-SNE visualization generated using the embedding vectors for different keywords or keyword combinations using the \texttt{VecVisualizer} module.}
    \label{fig:t_SNE}
\end{figure}

Figure~\ref{fig:t_SNE} provides a dimensionality reduced graphical representation of the relationships encoded in word embeddings between various properties associated with electrocatalysis.
Each point on the map corresponds to a specific keyword or keyword combination shown in the legend or its most similar terms based on the vector embedding.
Points that are close together in the visualization indicate properties or terms that frequently co-occur and therefore share context within the texts used. Different colors represent clusters of similar words related to a specific property.

Some key properties explored include ``electrocatalyst'', ``low overpotential'', ``high turnover frequency'', and ``favorable adsorption energy'', among others. The goal of this visualization is to discern how these properties interrelate and identify clusters that provide a comprehensive description of an electrocatalyst. By analyzing these clusters, users can compile a list of properties that encapsulate the defining features of an effective electrocatalyst as portrayed in the literature.

The t-SNE map, thus, serves as a companion tool~\cite{Maffettone2021}, aiding researchers in developing and refining their understanding and potentially directing them towards areas that warrant deeper investigation or properties or relationships that might have been overlooked.

\subsection{Exploring Material Similarities}

Understanding and describing complex relationship between materials themselves as well as their properties is typically a hard-learned skill through years of experience.
However, even experts with very good chemical intuition are usually not able to predict good candidate materials~\cite{D0SC04321D}.
One reason is that most compositionally complex materials are not just the sum of their individual components but the relationship between composition and resulting properties is typically non-linear.
\textsc{MatNexus} is a tool that allows for complex, high-dimensional analysis of relationships between composition and properties, not guided by intuition, but text data.

\begin{figure}[h]
    \centering
    \includegraphics[width=\textwidth]{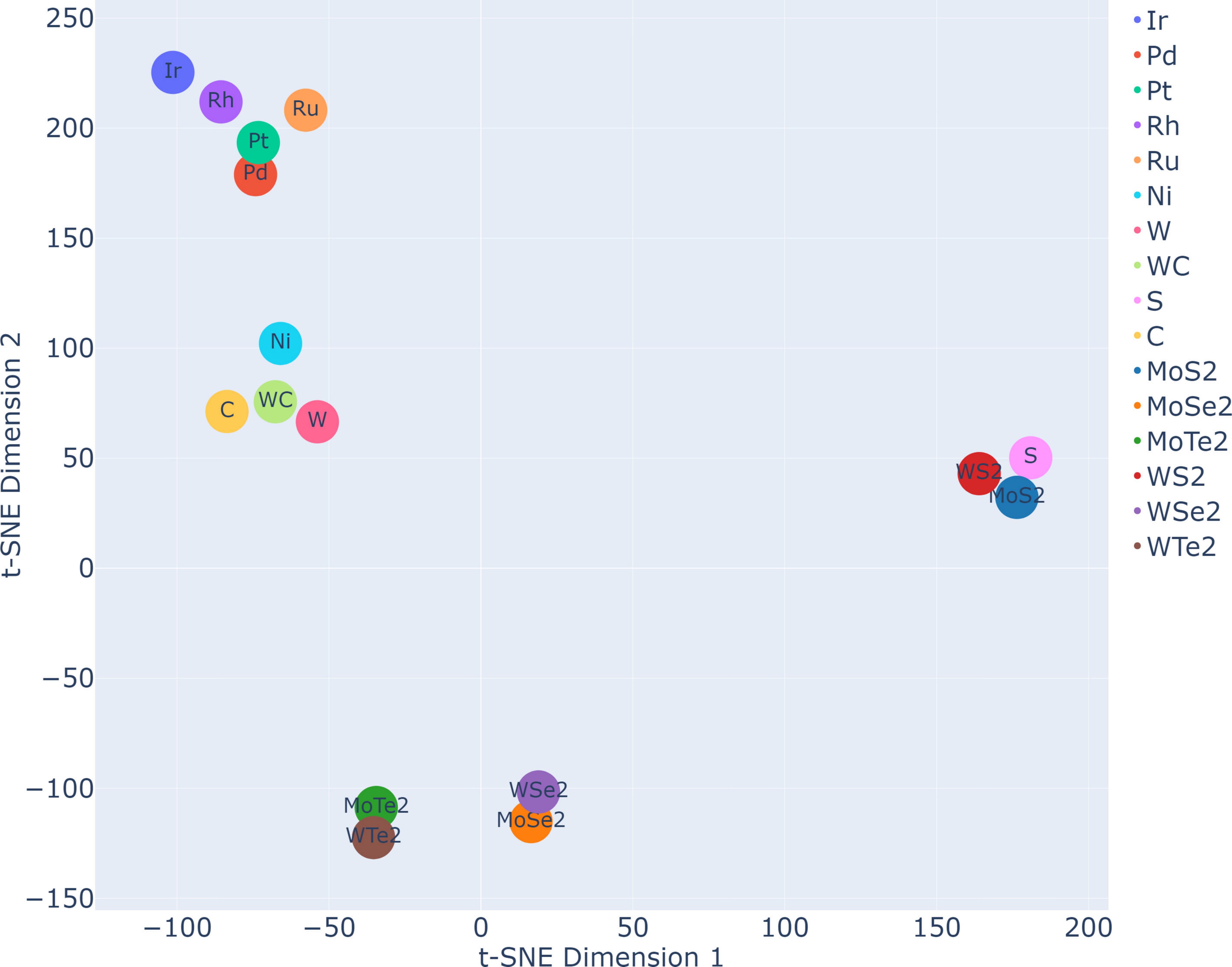}
    \caption{Visualization of material similarities as computed by \textsc{MatNexus} using dimension deduction technique of t-SNE.}
    \label{fig:material_similarity}
\end{figure}

As depicted in Figure~\ref{fig:material_similarity}, materials, originally represented as word embedding vectors in a high-dimensional space, are transformed for visualization purposes. Using dimensionality reduction techniques like t-SNE, the high-dimensional embeddings which capture the essence of the materials and their properties, are projected into a 2D space. In such a 2D representation, materials with similar properties or functionalities cluster together, allowing the user to visually discern and interpret their similarities. Notably:

\begin{itemize}
    \item Elements like Ni, Pd, Pt, Ru, Rh, and Ir form a tight cluster. This clustering suggests shared properties and potential applications as detailed in the literature.
    \item The chalcogenides, \( \text{WSe}_2 \), \( \text{WTe}_2 \), \( \text{MoSe}_2 \), and \( \text{MoTe}_2 \) are closely positioned, highlighting their similar characteristics.
    \item The remaining materials, namely WC, S, W, and C, showcase a different pattern. Particularly interesting is that WC is situated between W and C, and \( \text{WS}_2 \) lies closer to S than W, hinting at the composite nature of these materials and their relation to their constituent elements.
\end{itemize}

This graphical representation, generated using \textsc{MatNexus}'s visualizer tool enables researchers to understand the materials \textit{landscape}. Recognizing these clusters and interconnections allows for a deep exploration into specific groupings. By probing the shared properties within these clusters, researchers have the potential to unveil synergistic applications or innovative synthesis methods through interactive visualizations.

\section{Impact}

In all fields of science, the importance of effectively navigating the vast amount of literature cannot be understated~\cite{Pontius2020}. \textsc{MatNexus} is a tool to effectively access, process, and detect patterns and relationship based on textual data.
We show the example for electrocatalyst discovery but \textsc{MatNexus} is generally applicable.

\subsection{Addressing Existing Research Gaps}

Materials science, by its very nature, is based on data. Scientific literature, full with research findings and novel insights, is the primary source of new information~\cite{Himanen2019}.
However, the exponential growth in the volume of publications poses a challenge.
Researchers are now tasked with sifting through an overwhelming amount of information to extract relevant insights~\cite{Hill2016399}.
\textsc{MatNexus} is more than just a tool for accessing literature; it offers a structured methodology to gather, refine, analyze, and visually represent high-dimensional information in a compressed from scientific text sources.
However, the complexities in materials science are not merely about the vastness of data. The field of materials science is increasingly interwoven with other disciplines like chemistry, physics, and bioengineering~\cite{Khan2023491}. Such interdisciplinary overlaps have led to knowledge being dispersed across various domains, making it challenging to derive cohesive insights from scattered knowledge.
Unifying this information requires a tool that can integrate insights from diverse domains.
\textsc{MatNexus}, with its robust text processing capabilities, offers a unified view, ensuring researchers can interconnect insights and, more importantly, draw conclusions for future research directions not based on a vague notion of impression but based on data.

\subsection{Empowering Research Methodologies}

The traditional approach to literature review is often a linear path~\cite{SNYDER2019333}. Researchers might either dig deep into a narrow topic or end up with a broader yet superficial overview. \textsc{MatNexus} enables the transformation of the conventional approach.
By automating the data collection process and offering advanced analytical tools, it ensures that researchers no longer have to choose between depth and breadth.
They can quickly get an overview of a field or material and, with the same ease, delve deep into specific areas of interest by connecting to the original sources.

\subsection{Redefining Daily Research Workflows}

Incorporating \textsc{MatNexus} into a research routine can revolutionize conventional methodologies. Traditional literature reviews, indispensable as they are, remain time-intensive and often limiting~\cite{Kademani20131275}. The manual process of sifting through a myriad of publications, extracting salient data, and subsequently analyzing it has historically been a significant drain on a researcher's time. \textsc{MatNexus} streamlines this by automating much of the workflow, thus liberating researchers to direct their focus towards experimental endeavors and other research tasks.
The modern trend towards data-driven methodologies in materials science is evident in various pioneering works. For instance, \textit{Tshitoyan et al.}~\cite{Tshitoyan2019} capitalized on unsupervised word embeddings to extract latent knowledge from materials science literature. In another notable effort, \textit{Pei et al.}~\cite{Pei2023} employed text mining to propose the concept of ``context similarity'', thereby aiding in the selection of chemical elements for high-entropy alloys. Similarly, there have been dedicated initiatives to automate data pipelines for specific materials, such as superalloys~\cite{Wang2022}, or to extract inorganic materials synthesis recipes from an expansive corpus of literature~\cite{Kononova2019}.

While these efforts represent the current apex of text mining's potential in the domain of materials science, they often focus on specific procedures or tasks and do not provide a generally applicable open source tool for others. In contrast, \textsc{MatNexus} stands out as a widely applicable solution. It does not only provide tools for a singular aspect of the research process. Instead, it offers an integrated solution, allowing researchers to navigate the vast literature, extract and curate information, analyze data in high dimensions, and visualize intricate relationships, all within a singular ecosystem.

\subsection{Broader Implications and Widespread Adoption}

\textsc{MatNexus}'s potential extends far beyond the confines of materials science. Its basis in natural language processing and data analytics positions it as a versatile tool with applications that can potentially span a multitude of academic disciplines. While its current design is tailored for materials science, the underlying challenges it addresses – namely, (text) data overload, efficient data curation, and intricate data analysis – are prevalent in numerous academic fields.

Imagine the realm of biology, where researchers grapple with vast genomic datasets or the domain of medicine, where patient case studies and drug interaction data grow exponentially. Similarly, in social sciences, the sheer volume of qualitative data from surveys, interviews, and ethnographic studies can be overwhelming. In all these scenarios, a tool like \textsc{MatNexus}, appropriately adapted, can significantly enhance data-driven insights, streamline research processes, and catalyze novel discoveries.

\subsection{Future Commercial Prospects}

\textsc{MatNexus} holds significant promise beyond the academic realm, particularly for industries that heavily rely on research and development (R\&D). The ability to efficiently extract and analyze information from large volumes of scientific literature can markedly accelerate industry research cycles. This can, in turn, lead to accelerated innovation, reduced time-to-market, and more efficient product development processes.
Considering the commercial viability, a potential business model for \textsc{MatNexus} could involve offering a base version for general R\&D purposes, with premium, tailored modules catering to specific industry niches. These modules could be developed in collaboration with industry experts to ensure they address unique challenges faced by each sector.

However, venturing into the commercial space brings forth copyright considerations of the underlying text data. A feasible approach would involve primarily focusing on Open-access CC-BY licensed papers, ensuring compliance with existing copyright regulations. Collaborative arrangements with publishers or leveraging APIs that provide access to copyrighted content, under licensing agreements, could also be explored to expand the tool's dataset without infringing on copyrights.

\section{Conclusion}

\textsc{MatNexus} is an example of convergence of natural language processing and materials science research in one tool.
As the volume of scientific literature increases, the need for efficient navigation and interpretation tools becomes paramount. \textsc{MatNexus} addresses this by offering a streamlined, automated approach to data extraction and analysis. Its potential extends beyond academic realms, promising transformative impacts in industry-driven R\&D as well. We believe that the future of materials science will include tools like \textsc{MatNexus} to ensure comprehensive and efficient research, thereby enabling accelerated discoveries and innovations.

\section*{Acknowledgements}
The authors gratefully acknowledge the financial support provided by the China Scholarship Council (CSC, CSC number: 202208360048), which was instrumental in facilitating this research. 
\label{}

\bibliographystyle{elsarticle-num} 
\bibliography{Zhang_2023_MatNexus} 
\end{document}